\documentclass[aps,prl,onecolumn,superscriptaddress,10pt,floatfix,amsmath,amssymb]{revtex4-2}

\usepackage{graphicx} % Required for inserting images
\usepackage{dcolumn}% Align table columns on decimal point
\usepackage{xcolor}
\usepackage{bm}
\usepackage{physics}
\usepackage{lipsum}  
\usepackage{hyperref}
\usepackage[normalem]{ulem}
\graphicspath{{images/}}

\begin{document}

\title{Sum-frequency-based photon-number-resolving detector for telecom wavelengths}

\date{\today}

   \author{Silvia Cassina} \email{s.cassina@uninsubria.it}
   \affiliation{Como Lake Institute of Photonics, Dipartimento di Scienza e Alta Tecnologia, Università degli Studi dell'Insubria, Via Valleggio 11, I-22100 Como (Italy)}
   
   \author{Alex Pozzoli} 
   \affiliation{Como Lake Institute of Photonics, Dipartimento di Scienza e Alta Tecnologia, Università degli Studi dell'Insubria, Via Valleggio 11, I-22100 Como (Italy)}

     \author{Guglielmo Vesco}
   \affiliation{Dipartimento di Fisica, Politecnico di Milano, Polo Territoriale di Lecco, Via G. Previati 1/C,  
I-23900 Lecco (Italy)} 
   
   \author{Marco Lamperti}
   \affiliation{Como Lake Institute of Photonics, Dipartimento di Scienza e Alta Tecnologia, Università degli Studi dell'Insubria, Via Valleggio 11, I-22100 Como (Italy)} 
  
\author{Marco Marangoni}
   \affiliation{Dipartimento di Fisica, Politecnico di Milano, Polo Territoriale di Lecco, Via G. Previati 1/C,  
I-23900 Lecco (Italy)}   
   
   \author{Alessia Allevi} 
   \affiliation{Como Lake Institute of Photonics, Dipartimento di Scienza e Alta Tecnologia, Università degli Studi dell'Insubria, Via Valleggio 11, I-22100 Como (Italy)}

%%%%%%%%%%%%%%%%%%%%%%%%%%%%%%%%%%%%%%%%%%%%%%%%%%%%%%%%%%%%%%%
\begin{abstract}
The use of C-band wavelengths in the field of quantum communication has grown significantly, driving the need for versatile detection solutions, especially in the low intensity domain. Among the desirable features for such detectors, photon-number-resolving (PNR) capability is particularly valuable, since it can offer new possibilities for enhancing security of communication protocols. In this paper, we present the implementation of a receiver that combines low-cost PNR detectors with nonlinear optical interactions to achieve sensitivity at telecom wavelengths. Specifically, we use this receiver to characterize the Poissonian nature of a femtosecond source at 1.5 $\mu$m, produced via white light continuum generation followed by a single-stage amplification process.
The obtained results encourage the exploitation of such a detector in more complex schemes.
\end{abstract}

\maketitle

%%%%%%%%%%%%%%%%%%%%%%%%%%%%%%%%%%%%%%%%%%%%%%%%%%%%%%%%%%%%%%%%
\section{\label{sec:level1} Introduction}
The development of photon-number-resolving (PNR) detectors operating in the infrared spectrum has become a critical priority for advancing quantum communication technologies. Currently, telecommunication infrastructures benefit from optical fibers optimized for wavelengths around 1550 nm \cite{cariolaro}. However, the availability of efficient single-photon and PNR detectors in this spectral region is significantly limited compared to the visible spectral range.
Existing technologies, such as superconducting nanowire single-photon detectors, offer high detection efficiency and low dark counts, but require cryogenic cooling, increasing operational complexity and cost \cite{zhang,schmidt,gerrits}. Avalanche photodiodes represent another alternative \cite{campbell,dimario18}, but their photon-number resolution remains inadequate for advanced quantum protocols. Theoretical and experimental works demonstrated the potential of PNR detectors for applications such as secure quantum key distribution and quantum state tomography. For instance, in Refs.~\cite{cattaneo,notarnicola25} theoretical models have highlighted the critical role of photon-number resolution in improving the quantum channel capacity and the secret key generation rate, while the experimental results in Ref.~\cite{dimario19} have showcased the feasibility of PNR detectors in quantum optics. With respect to these applications, some of us have recently demonstrated the suitability of Silicon photomultipliers (SiPMs)\cite{OE24,PLA25}. Indeed, such detectors are compact, robust, low-cost, and have low-power requirements, thus being excellent candidates for quantum communication receivers. In this respect, we have shown that these detectors can be used to reveal nonclassical correlations of mesoscopic twin-beam states of light \cite{chesi,cassina,pozzoli} and to perform novel quantum communication protocols \cite{razzoli}. However, their response is limited to the visible and UV ranges.
To overcome this limitation, in this work we present a compact scheme based on cascaded nonlinear optical interactions, thus extending the operativity of SiPMs to the telecommunication range. In more detail, starting from a femtosecond laser source at 1030 nm, we produce and amplify a pulsed source at telecom wavelengths by means of a white light continuum (WLC) generation followed by an optical parametric amplifier (OPA). Then we convert the light at 1550 nm into light at 620 nm by collinear sum-frequency generation with the fundamental of the laser source. We characterize the statistical properties of the produced light in terms of the autocorrelation and cross-correlation functions \cite{pra12}. Specifically, we use the receiver operating at 620 nm to characterize the photon-number distribution of the amplified WLC and investigate its deviations from a Poissonian distribution.
The obtained results represent a first but fundamental step towards the implementation of a compact receiver operating at telecom wavelengths and characterized by an outstanding PNR capability. In particular, the ultimate goal is to implement a homodyne-like scheme based on SiPMs \cite{OE24}, in which the signal and local oscillator are produced at telecom wavelengths, while their detection is performed in the visible spectral range. The basic scheme includes two SiPMs, but more complex architectures, such as the HYNORE receiver in Ref.~\cite{notarnicola}, are also possible.   
%%%%%%%%%%%%%%%%%%%%%%%%%%%%%%%%%%%%%%%%%%%%%%%%%%%%%%%% 
\section{Materials and methods} \label{sec:methods} 
The proof-of-principle implementation of a PNR detector sensitive to telecom wavelengths relies on the production of a cascade of nonlinear interactions aimed at preparing, optimizing and properly tailoring the light source.
In view of highlighting the potential of the different steps introduced in the experimental setup, in the following we present a detailed description of each implemented process.
\subsection{Light generation and amplification at 1550 nm}
\begin{figure}
\centering
\includegraphics[width=8.9 cm]{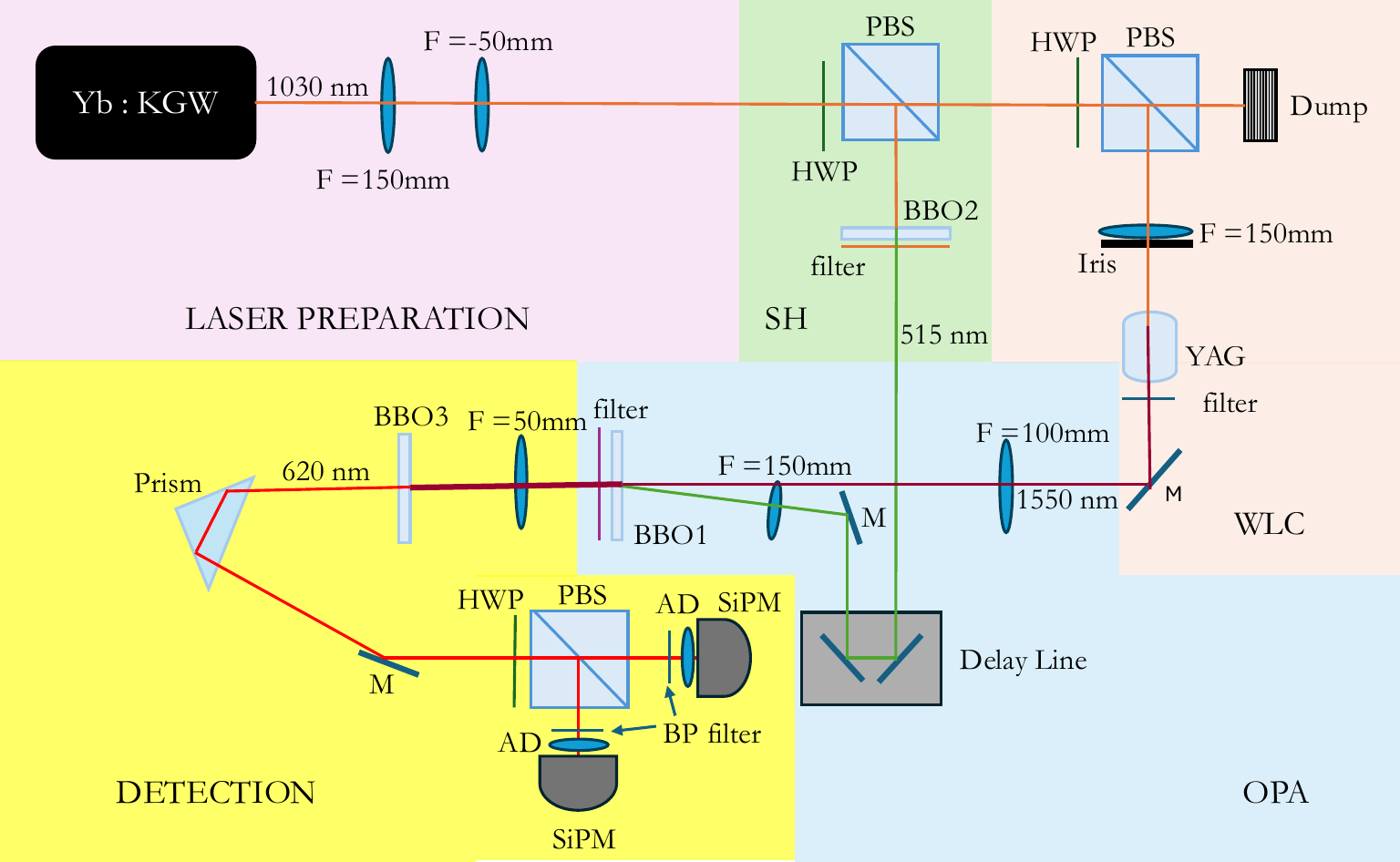}% Here is how to import EPS art
\caption{\label{setup} Sketch of the experimental setup used to produce both the beam at 1550 nm and the one at 620 nm. The colored boxes define the different steps of the setup. WLC: white light continuum; SH: second harmonic; OPA: optical parametric amplifier; F: convergent lens; HWP: half-waveplate; PBS: polarizing beam splitter; BBO1, BBO2, and BBO3: $\beta-$Barium-Borate crystals; YAG: yttrium aluminium garnet plate; M: mirror; BP: bandpass filter; AD: achromatic doublet; SiPM: Silicon photomultiplier. See the text for details.}
\end{figure}

\begin{figure}
\includegraphics[width=8 cm]{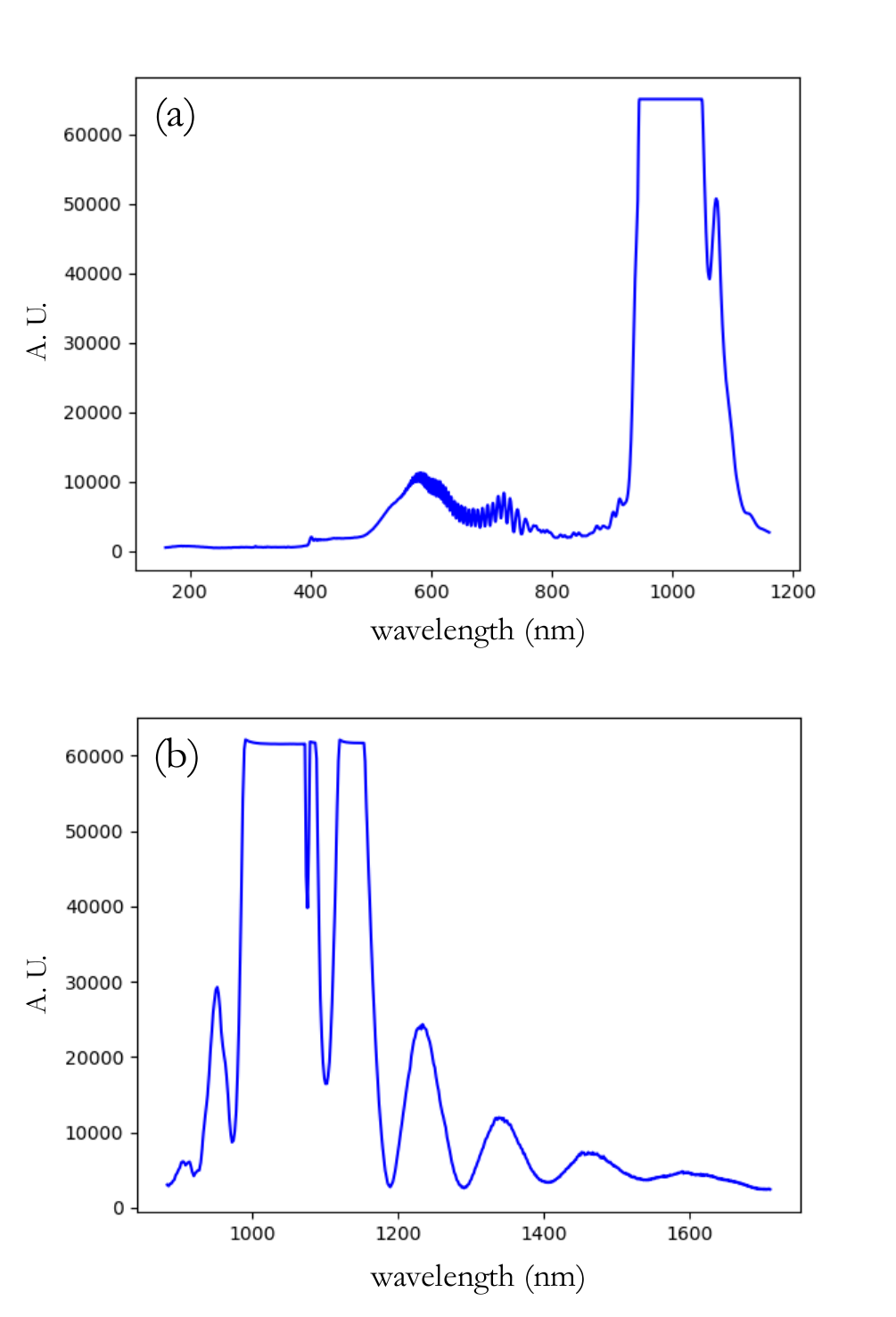}% Here is how to import EPS art
\caption{\label{SCspectrum} Spectrum of the white light continuum in the visible (panel (a)) and infrared region (panel (b)). In both panels the fundamental beam at $\sim 1030$ nm, whose intensity value exceeds the full-scale range, is also present.}
\end{figure}

As shown in Fig.~\ref{setup}, the 190 fs long pulses at 1030 nm of a Yb:KGW amplified laser operated at a repetition rate of 3 kHz are split into two parts by means of a sequence of half-waveplates and polarizing beam splitters (PBSs) in order to finely tune both their polarization and their energy. 
One portion of light is focused (the focal length of the employed convergent lens is $f = 150$ mm) into a 10 mm thick yttrium aluminium garnet (YAG) plate to produce a WLC source (see box WLC in the figure). By pumping the YAG plate with a pulse energy of 1.6 $\mu$J,  we obtain WLC light extending over the entire visible spectral range and spanning up to 1600 nm \cite{villa}, as shown in the two panels of Fig.~\ref{SCspectrum}. The spectra were obtained using two different fiber spectrometers: in panel (a) the measurements were taken with AvaSpec-ULS2048x64-EVO (Avantes), setting the integration time to 2.4 ms, while regarding panel (b) they were saved with the MiniSpectrometer TG-NIR: C11482GA (Hamamatsu Photonics), setting the integration time to 1 ms.
Even though the absolute values depend on the spectral sensitivity curve of the adopted devices, the power spectral density at 1550 nm results much lower than at shorter wavelengths. Since this could represent a limitation for the exploitation of this telecom source in communication protocols, where the signals are typically modulated and can undergo unavoidable losses, we decided to amplify the light by means of a single-pass OPA \cite{gucci}. As shown in the box called OPA in Fig.~\ref{setup}, the generated light at 1550 nm is focused by a 100 mm focal length lens and mixed in a $\beta$-Barium-Borate (BBO) crystal, cut for type-I phase matching (cut angle = 23.4$^{\circ}$, 3 mm long, called BBO1 in the figure), with the second harmonic (SH) (at 515 nm) of the laser by means of a slightly noncollinear interaction geometry to spatially separate the beams after the OPA.
The SH is generated by sending one of the infrared portions (with roughly 19 $\mu$J of energy per pulse) into a second BBO crystal cut for type-I phase matching (cut angle = 22.8$^{\circ}$, 3 mm long, called BBO2 in the box SH of the figure). SH is focused by a 150 mm focal-length lens in BBO1 and its temporal delay can be micrometrically varied in order to optimize the temporal overlap with the WLC pulses in the crystal. 
\begin{figure}
\includegraphics[width=8 cm]{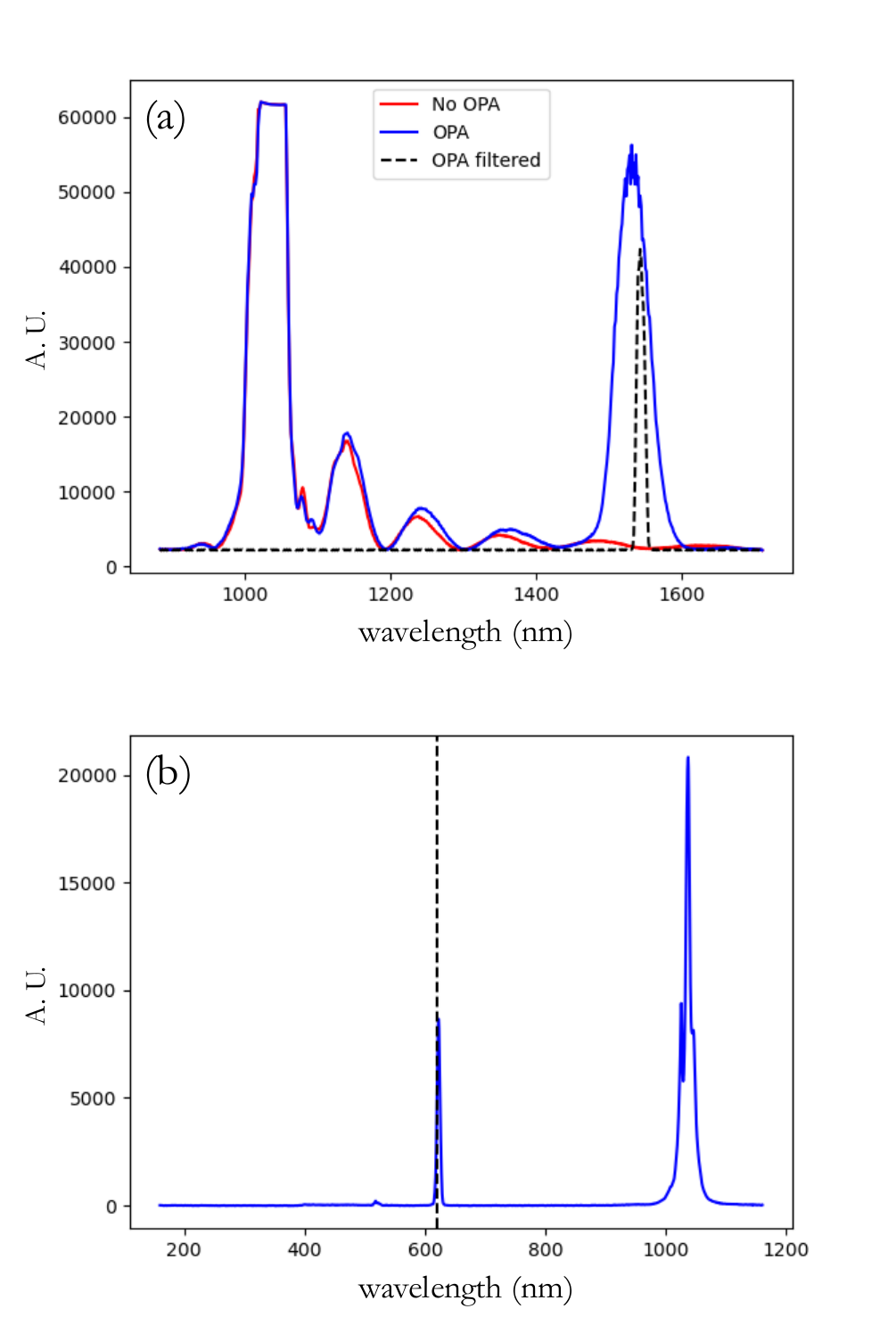}% Here is how to import EPS art
\caption{\label{1550spectrum} (a) Spectrum of the white light continuum in the infrared region in the presence (blue curve) and in the absence (red curve) of amplification. The dashed black curve corresponds to the amplified spectrum obtained by using a band-pass filter centered at 1550 nm in front of the spectrometer. (b) Spectrum of the light exiting BBO3, in which the generation of the sum frequency at 620 nm, highlighted by the black dashed line, is clearly evident.}
\end{figure}
The spectrum of the amplified light at 1550 nm is shown in Fig.~\ref{1550spectrum}(a), along with the spectrum obtained when SH light does not enter BBO1. The comparison highlights the importance of the amplification stage, which provides a gain of approximately $10^4$ in the region centered at 1550 nm, with a bandwidth of $\sim 50$ nm. Both spectra were obtained with the fiber spectrometer sensitive to the infrared spectral range. The spectrum of the amplified light with a band-pass filter centered at 1550 nm is also shown as dashed black curve to highlight the wavelength we are interested in.
%%%%%%%%%%%%%%%%%%%%%%%%%%%%%%%%%%%%%%%%%%%%%%%%
\subsection{Sum-frequency generation at 620 nm and detection}
\begin{figure}
\includegraphics[width=8 cm]{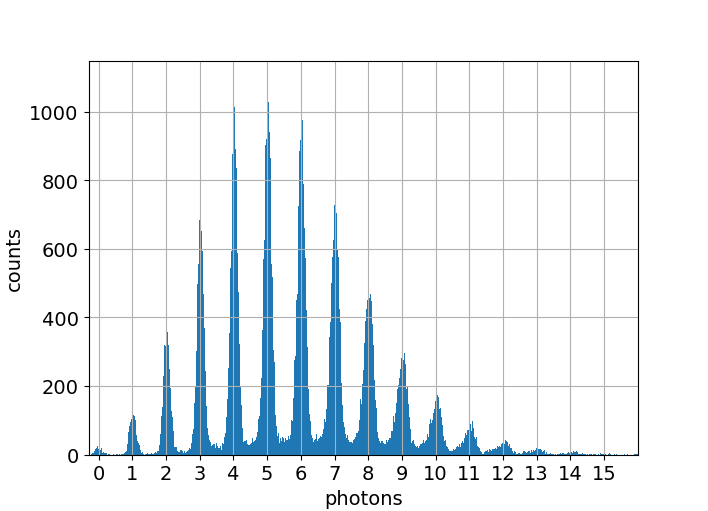}% Here is how to import EPS art
\caption{\label{PHS} Typical pulse height spectrum of a SiPM, from which it is evident its outstanding PNR capability.}
\end{figure}

As shown in the DETECTION box of Fig.~\ref{setup}, the amplified light at 1550 nm is directed to a further BBO crystal (cut angle = 23.4$^{\circ}$, 3 mm long, labelled BBO3 in the figure), where
%, where it has enough energy to produce its second harmonic at 780 nm. At variance with this nonlinear process, at a slightly different angle (the difference between the two tuning angles is equal to 0.1$^{\circ}$ inside the BBO) 
it interacts with the residual infrared light at 1030 nm traveling along the same optical path. This interaction leads to a sum-frequency generation process in a collinear interaction geometry. A typical spectrum of the generated light is shown in Fig.~\ref{1550spectrum}(b), which was obtained with the Avantes fiber spectrometer. The wavelength of the generated light matches the sensitivity range of the PNR detectors that are embedded in the hybrid receiver described in Ref.~\cite{OE24}.\\ 
As anticipated in the Introduction, these PNR detectors are SiPMs. They consist of matrices of cells operated in the Geiger-M$\ddot{\rm u}$ller regime with a common output. The model used (mod. MPPC S13360–1350CS, Hamamatsu Photonics) in this proof-of-principle work has a sensitivity peak in the blue-green region. However, there are also other models more suitable for the detection in the red spectral range. SiPMs are endowed with an outstanding PNR capability even if they are operated at room temperature, as shown in Fig.~\ref{PHS}, which reproduces a typical pulse height spectrum. 
The two detectors we used to perform the characterization of the generated light are amplified by a fast amplification stage, then the output signals are integrated over a 10 ns long gate to reduce the contributions of dark counts and optical cross-talk, which generally affect SiPMs\cite{scirep19,cassina}. As to dark counts, by performing an experimental dark measurement of $10^5$ samples, we experimentally obtained a value of about 0.3$\%$ of dark counts for both the employed detectors. Finally, the analog signals are digitized and acquired.
Even though the hybrid scheme in Ref.~\cite{OE24} involves an interferometric scheme and a local oscillator, like in the case of standard optical homodyne detection schemes \cite{lvovsky}, the preliminary characterization performed here involves the two SiPMs in a different configuration. Indeed, since we are mainly interested in investigating the coherence properties of light generated at 620 nm with the current setup, we used a scheme for the measurements of $g^{(2)}$ autocorrelation and $g^{(1,1)}$ cross-correlation functions. As shown in Fig.~\ref{setup}, at the output of BBO3 we place a prism to separate the different wavelengths exiting the crystal, i.e. 620 nm, 1030 nm, and 1550 nm. The red part is then divided at a PBS preceded by a half waveplate to finely tune the balancing between the two output arms of the PBS. Moreover, each output is focused by an achromatic doublet into a multi-mode fiber with 1 mm core diameter and delivered to a SiPM.
\subsection{Theoretical framework}
The statistical characterization of the light produced at 620 nm can be achieved by considering either the first moments of the photon-number distribution or the photon-number distribution itself.
Indeed, starting from a coherent beam, we expect that the sequence of nonlinear interactions we are realizing does not change its statistical properties \cite{debethune,chmela}. In that case, the statistical distribution of detected photons, to which we have direct access \cite{arimondo,APL}, is described by a Poissonian distribution \cite{mandel}:
\begin{equation}\label{poiss}
    p(m) = \frac{\langle m \rangle^m}{m!} \exp \left(-{\langle m \rangle} \right),
\end{equation}
in which $\langle m \rangle$ is the mean number of detected photons.
Moreover, it could be useful to characterize the generated light by considering the autocorrelation and cross-correlation (at the PBS) functions, since their value can easily reveal the statistical features of the generated light \cite{avenhaus,vogel,perina}. 
According to the definition given by Glauber \cite{glauber}, the second-order autocorrelation function is defined as 
\begin{equation}\label{g2photon}
    g^{(2)}(n) = \frac{\langle :n^2: \rangle}{\langle n \rangle^2} = \frac{\langle n^2 \rangle}{\langle n \rangle^2} - \frac{1}{\langle n \rangle} = 1 + \frac{F(n) -1}{\langle n \rangle},
\end{equation}
where $: \cdot :$ refers to the normal ordering operation and $F(n)$ is the Fano factor $F(n) = \sigma^2(n)/\langle n \rangle$, in which $\sigma^2(n)$ is the variance of the distribution. In the case of Poissonian light, since all the moments of the distribution are equal, $F(n) = 1$.
In terms of detected photons, where $\langle m \rangle = \eta \langle n \rangle$, $F(m) = \eta F(n) + (1- \eta)$, in which $\eta$ represents the quantum efficiency. Hence, an expression analogous to that in Eq.~(\ref{g2photon}) can be written for detected photons as
\begin{equation}\label{g2detphoton}
    g^{(2)}(m) = \frac{\langle m^2 \rangle}{\langle m \rangle^2} = g^{(2)}(n) + \frac{1}{\langle m \rangle} = 1 + \frac{F(m) -1}{\langle m \rangle} + \frac{1}{\langle m \rangle}.
\end{equation}
In the case of Poissonian light we have
\begin{equation}\label{g2Poiss}
    g^{(2)}(m) = 1 + \frac{1}{\langle m \rangle}.
\end{equation}
By dividing a Poissonian light at a PBS, it is possible to calculate the shot-by-shot photon-number correlations or the cross-correlation function, $g^{(1,1)}$.
As shown in Ref.~\cite{pra12}, in terms of detected photons it is defined as
\begin{equation}\label{g11}
    g^{(1,1)}(m) = \frac{\langle m_1 m_2 \rangle}{\langle m_1 \rangle \langle m_2 \rangle},
\end{equation}
in which $m_1$ and $m_2$ are the numbers of detected photons at the two outputs of the PBS. Since in the case of Poissonian light $\langle m_1 m_2 \rangle = \langle m_1 \rangle \langle m_2 \rangle$, Eq.~(\ref{g11}) reduces to
\begin{equation}\label{g11Poiss}
    g^{(1,1)}(m) = 1.
\end{equation}
%%%%%%%%%%%%%%%%%%%%%%%%%%%%%%%%%
\section{Results}\label{sec:results}
Since this work is aimed at demonstrating that the light generated in the mesoscopic intensity regime through the sequence of nonlinear interactions is statistically stable and described by a Poissonian distribution, the first kind of investigation we decided to carry out concerns the statistical properties of the generated light as a function of the power, $P$, of the fundamental beam (at 1030 nm) used to produce WLC.
Indeed, while many characterizations of the coherence of WLC light have been performed in the spectral and spatial frameworks \cite{majus, vandewalle, dubietis, halder}, to the best of our knowledge a systematic analysis in terms of photon-number statistics still lacks.
\subsection{Critical power for WLC generation}
\begin{figure}
\includegraphics[width=8 cm]{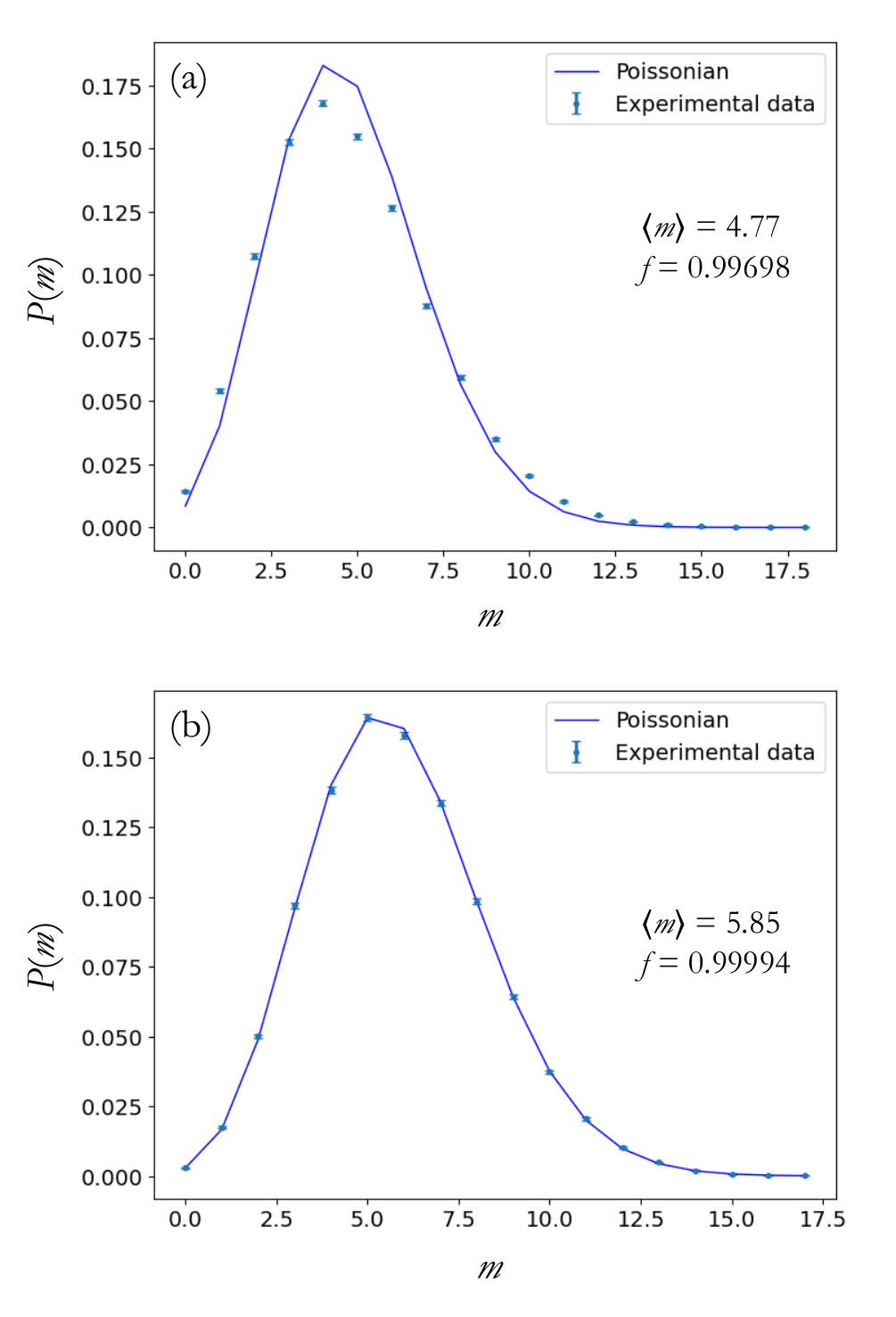}% Here is how to import EPS art
\caption{\label{statVSpower} (a) Reconstructed detected-photon distribution of the up-converted beam at 620 nm when $P \sim 4P_{\rm cr}$ and $\langle m \rangle = 4.77$. (b) The same as in (a) when $P \sim 6P_{\rm cr}$ and $\langle m \rangle = 5.85$. In both panels dots + error bars correspond to experimental data, while the solid curve is the theoretical expectation according to Poisson distribution in Eq.~(\ref{poiss}). The fidelity values to Poissonian distributions are $f=0.99698$ in panel (a) and $f=0.99994$ in panel (b).}
\end{figure}

\begin{figure}
\includegraphics[width=8 cm]{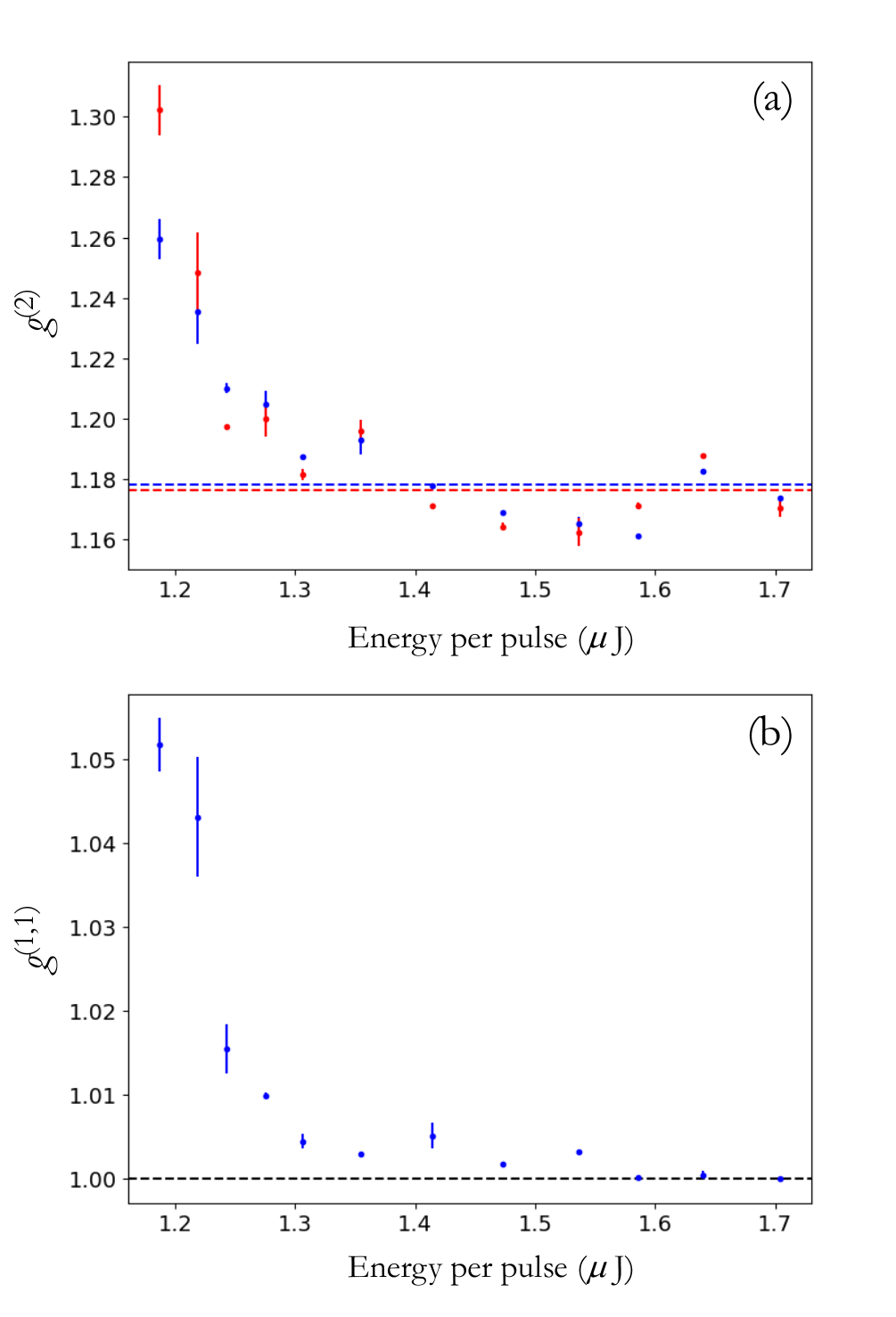}% Here is how to import EPS art
\caption{\label{gVSP} (a) $g^{(2)}$ autocorrelation function as a function of the energy per pulse of the fundamental beam at 1030 nm. Colored dots + error bars: experimental data corresponding to the two SiPMs; colored dashed lines: expected values in the case of Poissonian statistics with mean value roughly equal to $\langle m \rangle = 5.6$. (b) $g^{(1,1)}$ cross-correlation function as a function of the power of the fundamental beam at 1030 nm. The black dashed line is the expected value in the case of Poissonian statistics.}
\end{figure}

In the two panels of Fig.~\ref{statVSpower} we show the photon-number distributions of detected photons at two different values of $P$: $\sim 4$ times the critical power in panel (a) and $\sim 6$ times the critical power in panel (b). As to the critical power for self-focusing, we have considered Marburger's formula \cite{marburger}, which reads as
\begin{equation}\label{Pcr}
    P_{\rm cr} = \frac{3.72 \cdot \lambda_0^2}{8 \pi n_0 n_2},
\end{equation}
in which the numerical coefficient, 3.72, is valid for a cylindrically symmetric Gaussian beam.
Assuming $\lambda_0 = 1030$ nm, $n_0 = 1.82$ and $n_2 = 6.13 \cdot 10^{-20}$ m$^2$/W for the YAG material we used, the critical power happens to be $\sim 1.4$ MW, corresponding to an energy of 0.3 $\mu$J per pulse. 
In both panels we report as a solid line the Poissonian distribution calculated using the measured mean number of detected photons. The overlap between this distribution and the experimental data is rather poor in panel (a), indicating that the reconstructed distribution deviates from a Poissonian form. In contrast, panel (b) shows a much better agreement, suggesting a closer match to the expected Poissonian behavior. To quantify the discrepancy between theory and experiment, we consider the fidelity parameter, which is defined as $f = \sum_{m = 0}^{\bar{m}} \sqrt{p_{\rm th}(m) p_{\rm exp}(m)}$, in which $p_{\rm th}(m)$ and $p_{\rm exp}(m)$ are the theoretical and experimental distributions, respectively, and the sum extends up to the maximum number of detected photons, $\bar{m}$, above which both $p_{\rm th}(m)$ and $p_{\rm exp}(m)$ become negligible. Indeed, we get $f=0.99698$ for panel (a) and $f=0.99994$ for panel (b).
To systematically study how the power of the fundamental beam affects the stability of the WLC and thus the Poissonian character of the light at 620 nm, we consider both the autocorrelation function $g^{(2)}$ and the cross-correlation function $g^{(1,1)}$ shown in the two panels of Fig.~\ref{gVSP}, respectively, for roughly the same mean value $\langle m \rangle$, obtained by means of a neutral density filter placed beyond the prism in the DETECTION box. 
We note that there is a decreasing trend of $g^{(2)}$ at increasing values of the energy per pulse of the fundamental beam, and thus of its power. In particular, $g^{(2)}$ values larger than 1.18, corresponding to the expected value in Eq.~(\ref{g2Poiss}) for $\langle m \rangle = 5.6$, prove that values of $P$ close to $P_{\rm cr}$ determine fluctuations in the number of photons. The same kind of information can be extracted by looking at the behavior of $g^{(1,1)}$, which is equal to 1 only for values of the energy per pulse larger than 1.58 $\mu$J.
%%%%%%%%%%%%%%%%%%%%%%%%%%%%%%%%%%%%%%%%%%%%%%%%
\subsection{Poissonian character of the upconverted field}
\begin{figure}
\includegraphics[width=8 cm]{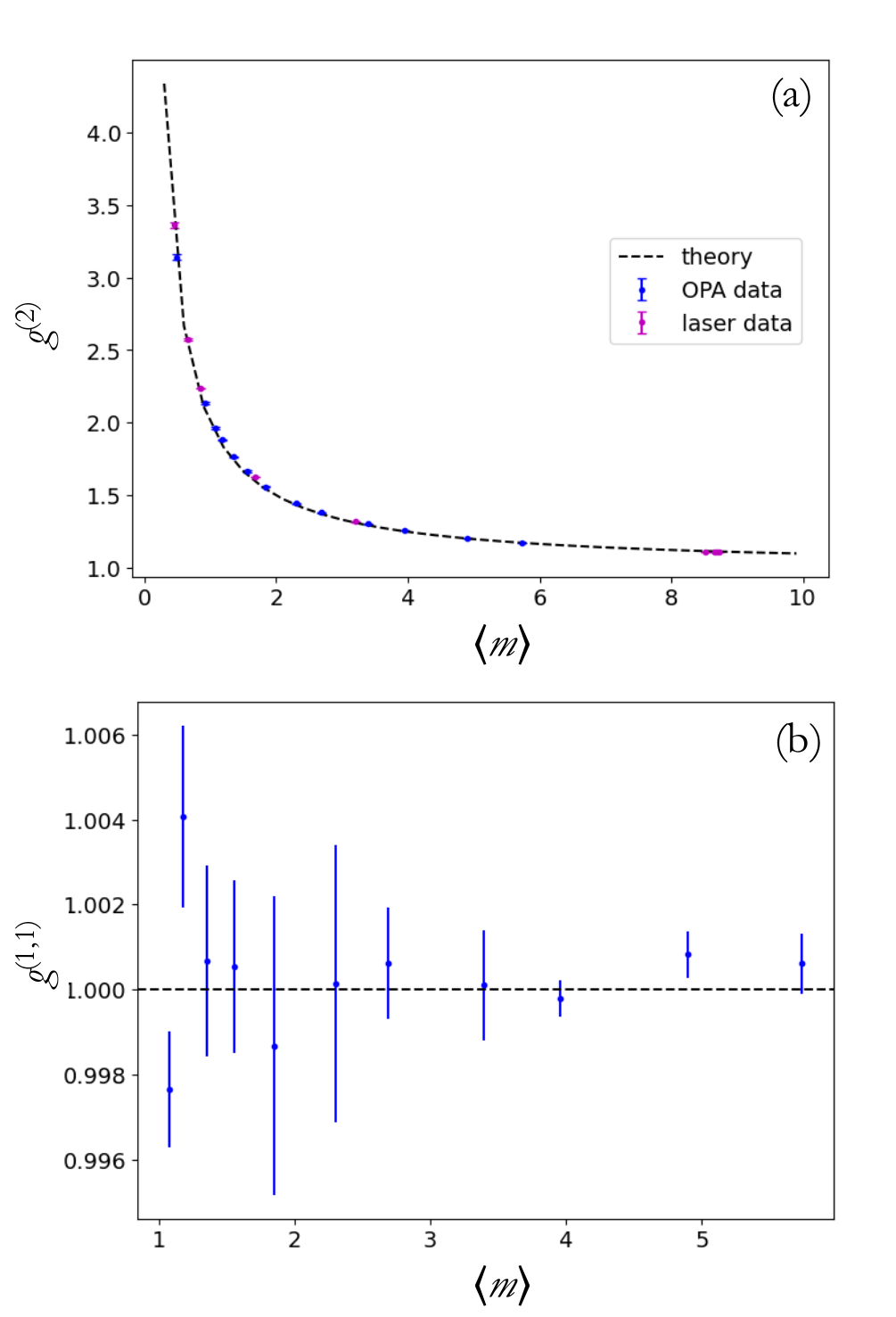}% Here is how to import EPS art
\caption{\label{gVSmean} (a) $g^{(2)}$ autocorrelation function as a function of the mean number of photons detected by the SiPM in the transmitted arm of the PBS in the DETECTION box of Fig.~\ref{setup}. Blue dots + error bars: experimental data corresponding to the beam at 620 nm; magenta dots + error bars: experimental data corresponding to the second harmonic of the laser; black dashed line: theoretical expectation according to Eq.~(\ref{g2Poiss}). (b) $g^{(1,1)}$ cross-correlation function as a function of the mean number of photons detected by the SiPM in the transmitted arm of the PBS in the box DETECTION. The black dashed line is theoretical expectation according to Eq.~(\ref{g11Poiss}).}
\end{figure}

The study reported in the previous Section, which examined the photon-number statistics as a function of the 1030 nm pulse power used for WLC generation, identified 1.7 µJ as a safe energy value to prevent deviations from Poissonian statistics.
In this Section, by keeping the fundamental pulse energy for WLC fixed at this value, we studied if the Poissonian character is preserved by varying the mean number of photons of the beam at 620 nm to be sure that the sum-frequency generation process does not introduce any additional noise. For this purpose a continuously variable neutral density filter was used to finely adjust its energy per pulse.
\begin{figure}
\includegraphics[width=8 cm]{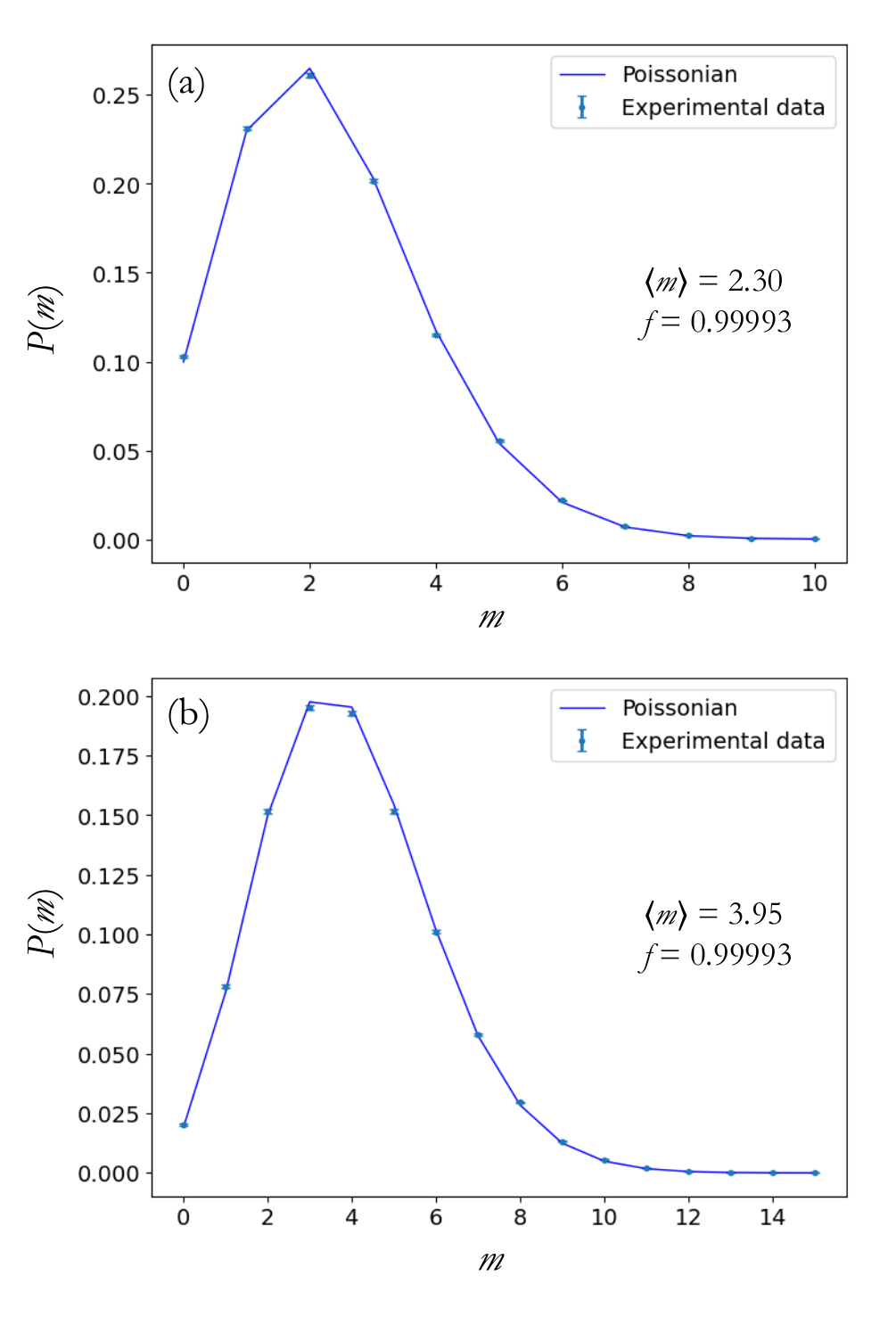}% Here is how to import EPS art
\caption{\label{statVSmean} (a) Reconstructed detected-photon distribution of the up-converted beam at 620 nm for $\langle m \rangle = 2.30$. (b) The same as in (a) for $\langle m \rangle = 3.95$. In both panels dots + error bars correspond to experimental data, while the solid curve is the theoretical expectation according to Poisson distribution in Eq.~(\ref{poiss}). The fidelity values to Poissonian distributions are $f=0.99993$ in both cases.}
\end{figure}
Figure~\ref{statVSmean} shows the reconstructed detected-photon distribution at different mean values together with the corresponding Poissonian distributions. High values of fidelity ($f > 0.9999$) prove that the generation of WLC is reasonably stable in the investigated conditions.
To better emphasize the Poissonian nature of light, we show in Fig.~\ref{gVSmean} 
the values of $g^{(2)}$ and $g^{(1,1)}$ as functions of the mean number of photons detected in the transmitted arm of the PBS placed in the box DETECTION of Fig.~\ref{setup}. To prove that the obtained values correspond to a reliable and stable condition, in panel (a) of the figure we also show the results achieved by sending to SiPMs the built-in second-harmonic beam (at 515 nm) of the laser, whose photon-number distribution is described by a Poisson distribution \cite{OE24}. For the SH and the up-converted beam at 620 nm the measured values of the autocorrelation function are in good agreement and they are superimposed by the same theoretical curve described by Eq.~(\ref{g2Poiss}), suggesting therefore that the up-converted beam follows a Poisson distribution.
Moreover, in panel (b) of Fig.~\ref{gVSmean} the calculated values of $g^{(1,1)}$, namely the cross-correlation between the numbers of photons detected at the two PBS outputs, randomly fluctuate around 1, thus indicating that the statistical distribution is Poissonian at any mean values of detected photons.  
%%%%%%%%%%%%%%%%%%%%%%%%%%%%%%%%%%%%%%
\section{Discussion}\label{discussion}
The results obtained as a function of the power of the fundamental beam prove that there is a strict dependence of the statistical properties of the up-converted beam on the stability of the WLC. In fact, appreciable deviations from the Poissonian character of light produced at 620 nm are observed even when the investigated power values are above $P_{\rm cr}$. This is an interesting result itself, as to the best of our knowledge, photon-number properties of WLC are not a deep-investigated subject. 
Further investigations on this topic could be addressed using different nonlinear materials, such as potassium gadolinium tungstate (KGW), which can produce a WLC more extended in the infrared spectral range \cite{marciulionyte}. In particular, it would be interesting to study the stability of WLC as a function of the power and compare the results with those obtained with the YAG plate. In addition, it might be useful to analyze the dependence of the statistical properties on the beam waist size and on the beam waist position with respect to the input face of the nonlinear crystal, as these represent two additional critical parameters alongside power.\\  
The investigation carried out in this work represents a crucial point to be fixed in view of the application of the generated light in communication protocols, where the statistical properties are suitably modulated \cite{grosshans}.\\ 
Moreover, the study of the photon-number distribution by varying the mean number of photons proves that the source produced at 620 nm does not introduce any noise, thus fully respecting the Poissonian character at any intensity values that can be explored by means of SiPMs. This supports the fact that the chosen value of the power of the fundamental beam is sufficient to guarantee the robustness of the source. Even the comparison with a native Poissonian light proves this statement. 
This represents a crucial step towards the further development of the receiver. One of the improvements is the realization of a sum-frequency process that uses two different paths for the light at 1030 nm and that at 1550 nm in order to suitably manipulate and modulate the beam only at the telecom wavelengths.
%%%%%%%%%%%%%%%%%%%%%%%%%%%%%%%%%%%%%%
\section{Conclusions}\label{conclusions}
In this work we have presented the proof-of-principle implementation of a receiver based on SiPMs and nonlinear optical interactions. The detector has been exploited to characterize a femtosecond laser source at 1550 nm produced by means of an WLC process followed by an OPA. The detector has enabled the investigation of the statistical nature of WLC light as a function of the power of the fundamental beam at 1030 nm. The achievement of a stable condition has been investigated as a function of the mean number of photons detected at 620 nm. The use of two SiPMs has allowed us to perform all these studies both in terms of the autocorrelation and cross-correlation functions.
The agreement of the results with the theoretical expectations opens new perspectives in the use of such a detector in more complex schemes and for the investigation of more useful optical states, such as amplitude-modulated light beams.
%%%%%%%%%%%%%%%%%%%%%%%%%%%%%%%%%%%%%%%%%%%%%%%%%%%%%%%%%%%%%%%%

\section*{Acknowledgments}
M.L. and A.A. acknowledge support from Grant No. PNRR D.D.M.M. 737/2021, S.C. that by PNRR D.D.M.M. 351/2022. 
Scientific support from CRIETT centre of University of Insubria (instrument code: MAC27) is greatly acknowledged.
The authors acknowledge Dr. Pietro Anzini (University of Insubria) for the loan of the fiber spectrometer MiniSpectrometer TG-NIR: C11482GA.

\section*{Author declaration}
\subsection*{Conflict of interest}
The authors have no conflicts to disclose.

\section*{Data Availability Statement}
The datasets generated and analyzed during the current study are available from the corresponding authors on reasonable request.

\end{document}